\documentclass[journal]{IEEEtran}
\usepackage{graphicx}

\usepackage[modulo]{lineno}

\begin{document}
\title{Irradiation induced effects in the FE-I4 front-end chip of the ATLAS IBL detector}

\author{
Alessandro La Rosa\\ on behalf of ATLAS collaboration
\thanks{Alessandro La Rosa is with the Max-Planck-Institut f\"ur Physik (Werner-Heisenberg-Institut), F\"ohringer Ring 6, D-80805 M\"unchen, Germany (telephone: +41-22-76-63600 , e-mail: alessandro.larosa@cern.ch).}
}%

\maketitle
\pagestyle{empty}
\thispagestyle{empty}

\begin{abstract}
The ATLAS Insertable B-Layer (IBL) detector was installed into the ATLAS experiment in 2014 and
has been in operation since 2015.
During the first year of IBL data taking an increase of the low voltage currents associated with the FE-I4
front-end chip was observed and this increase was traced back to the radiation damage in the chip. The
dependence of the current on the total-ionising dose and temperature has been tested with X-ray and proton
irradiations and will be presented in this paper together with the detector operation guidelines.
\end{abstract}


\section{Introduction}

\IEEEPARstart{A}{TLAS}~\cite{ATLAS} is a general-purpose experiment operating at the Large Hadron Collider (LHC) at CERN. The ATLAS detector was designed to be sensitive to a wide range of physics signatures to fully exploit the physics potential of the LHC at a nominal luminosity of $10^{34}$\,cm$^{-2}$s$^{-1}$. As most of the final states of collisions in the ATLAS experiment include charged particles, an excellent tracking system is essential.

The ATLAS Insertable B-Layer (IBL)~\cite{IBLTDR} is the innermost layer of the ATLAS pixel detector~\cite{PIXEL}. It is one of the major upgrades to the ATLAS experiment carried out during the long shutdown of the LHC in 2013--2014.

The IBL detector is the additional fourth pixel layer that was built around the new beryllium beam pipe and then inserted inside the Pixel detector in the core of the ATLAS detector.  It consists of 14 carbon fibre staves instrumented along 64\,cm, 2\,cm wide, and tilted in $\phi$ by 14$^{o}$  surrounding the beam-pipe at a mean radius of 33\,mm  from the beam axis and providing a pseudo-rapidity coverage of $\pm$\,3. Each stave, with integrated CO$_{2}$ cooling, is equipped with 32 front-end chips (FE-I4~\cite{FEI4}) bump bonded to silicon sensors.

The FE-I4 chip is designed in 130\,nm CMOS technology which features an array of 80\,x\,336 pixels with a pixel size of 50\,x\,250\,$\mu$m$^{2}$.
Each pixel contains an independent, free running amplification stage with adjustable shaping, followed by a discriminator with independently adjustable threshold. The FE-I4 keeps track of the time-over-threshold (ToT) of each discriminator with 4-bit resolution, in counts of an external supplied clock of 40\,MHz frequency.
The FE-I4 operates by feeding the common power supply to analog signal amplifiers and digital signal-process circuits, referred to as the low-voltage (LV) power supply and the clock input.

The IBL detector is designed to be operational until the end of the LHC Run~3, where the total integrated luminosity is expected to reach 
$300~{\rm fb^{-1}}$. The detector components are qualified to work up to 250\,Mrad of total ionising dose (TID).

During the first year of the IBL operation in 2015 a significant increase of the LV current of the front-end chip and the detuning of its parameters (threshold and time-over-threshold) have been observed in relation to the received TID.
In this paper, the TID effects in the FE-I4 chip are reported based on studies performed in the laboratory using X-ray and proton irradiation sources for various temperature and irradiation intensity conditions. Based on these results, an operation guideline of the IBL detector is presented.

\section{Observations}

\begin{figure}[h!]
\centering
\includegraphics[width=3.5in]{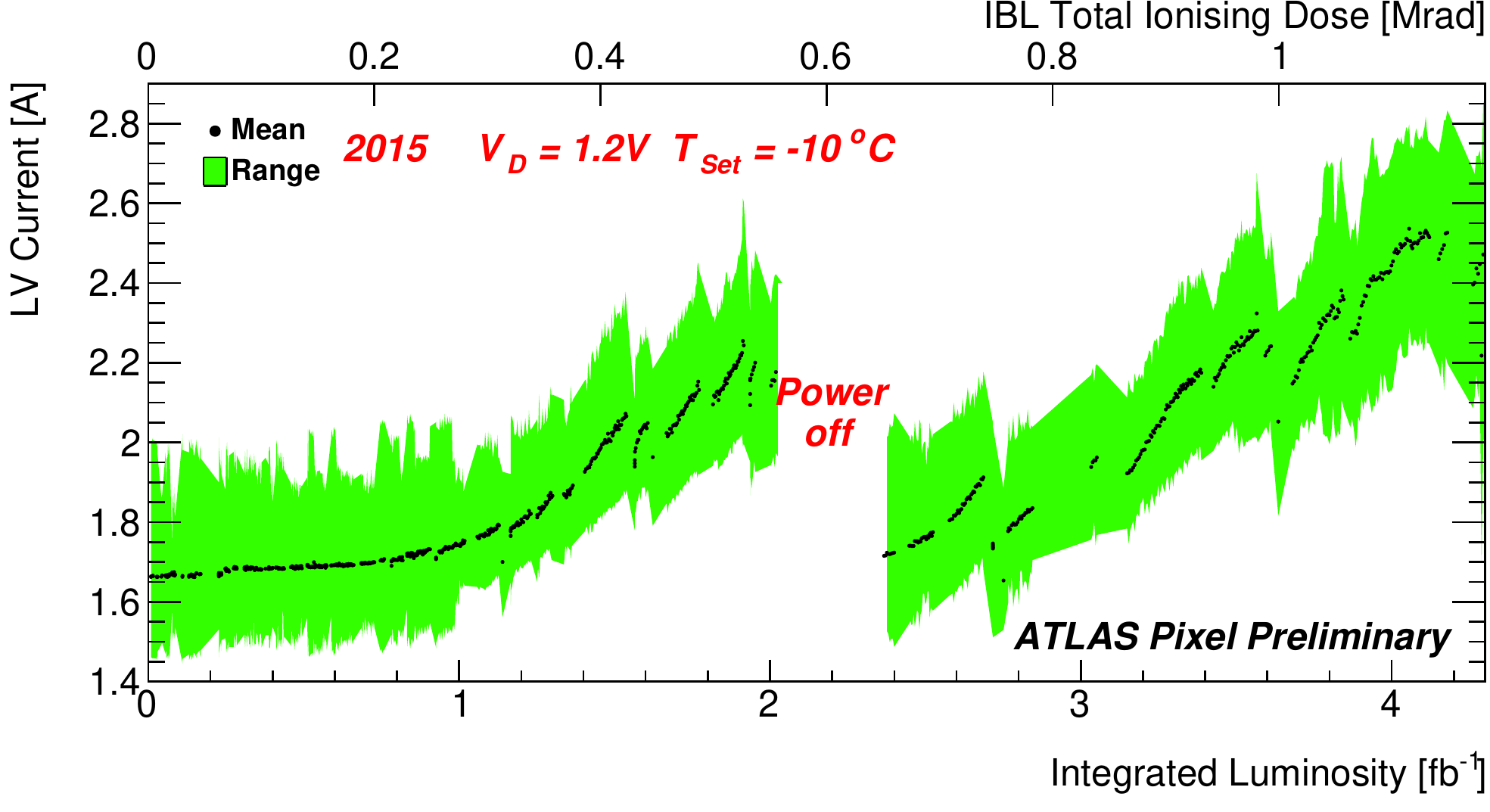}
\caption{Mean low voltage (LV) current in IBL FE-I4 chips during stable beam as a function of integrated luminosity and total ionising dose (TID). In the period from September to November 2015 the IBL detector was switched off during one LHC fill (due to safety concerns in early October 2015). The mean LV currents are averaged for all modules across 100 luminosity blocks and there is no obvious dependence of LV current on module group position. The TID is calculated from integrated luminosity~\cite{IBLLVcurrent}.}
\label{fig:IBL_LVdrift2015}
\end{figure}

During the operation of the IBL detector, the LV current of the FE-I4 chip was stable at a value of 1.6--1.7\,A (for a four-chip unit) until the middle of September 2015.  Then, the current started to rise up significantly (see Figure~\ref{fig:IBL_LVdrift2015}), and the change of the current during September to November 2015 was more than 0.2\,A even within a single LHC fill, depending on the luminosity and the duration of the fill.

\begin{figure}[h!]
\centering
\includegraphics[width=3.5in]{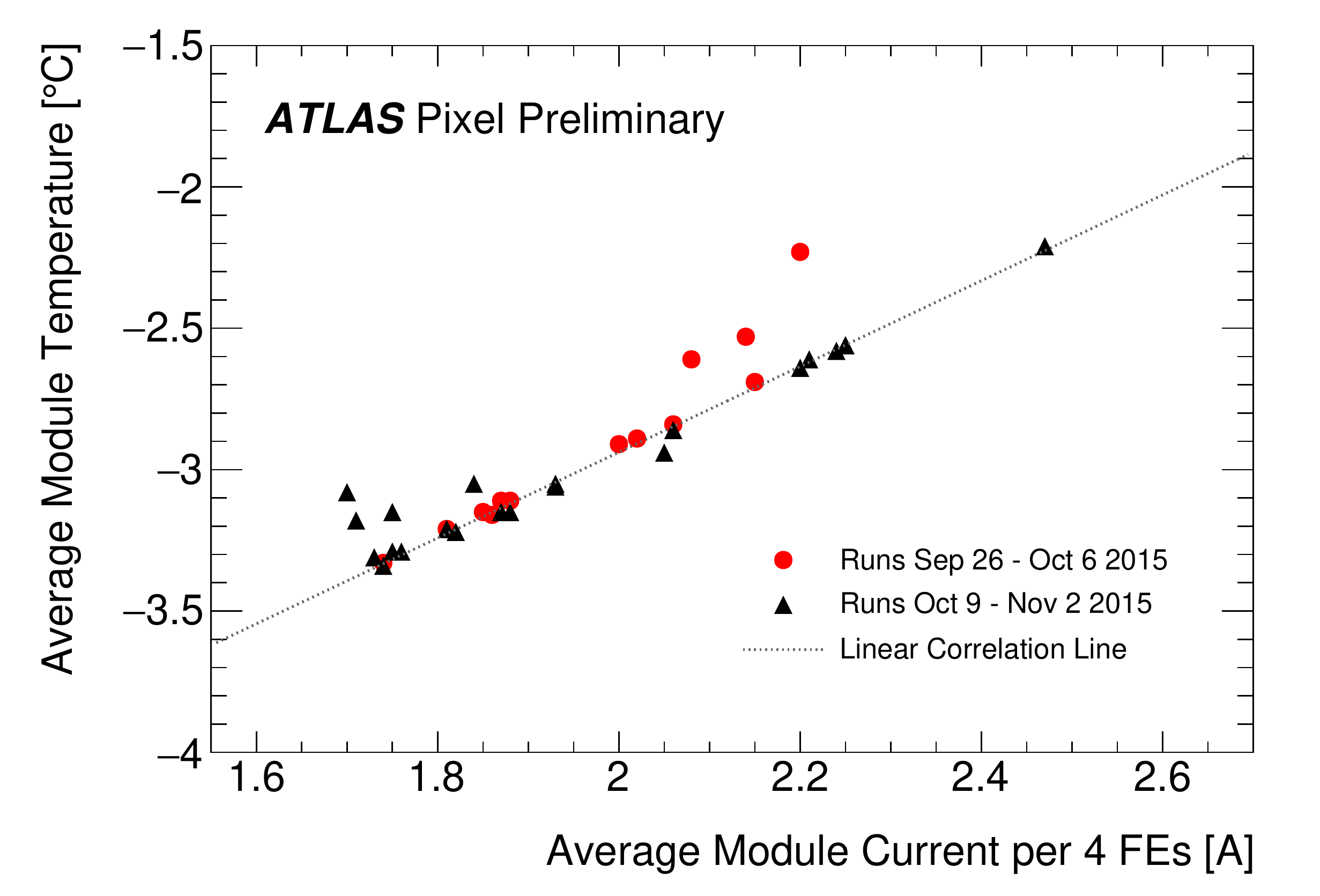}
\caption{Performance of the IBL modules during high luminosity proton-proton collision runs from September to November 2015, separated into the periods before (red circles) and after (black triangles) the long power-off on October 6. The data are displayed as a function of the average module current per 4 front-ends of the IBL and compared to a linear dependence. The average module temperature is shown~\cite{CurrentTemp}.}
\label{fig:correlation_T_I_2015}
\end{figure}

With the increase of the LV current, the temperature of IBL modules also changes (Figure~\ref{fig:correlation_T_I_2015}). The change of the thermo-mechanical condition of the IBL resulted in the change of the IBL distortion magnitude, and a clear relation between the module temperature and the distortion magnitude was observed~\cite{DistortionNote}.

\begin{figure}[h!]
\centering
\includegraphics[width=3.5in]{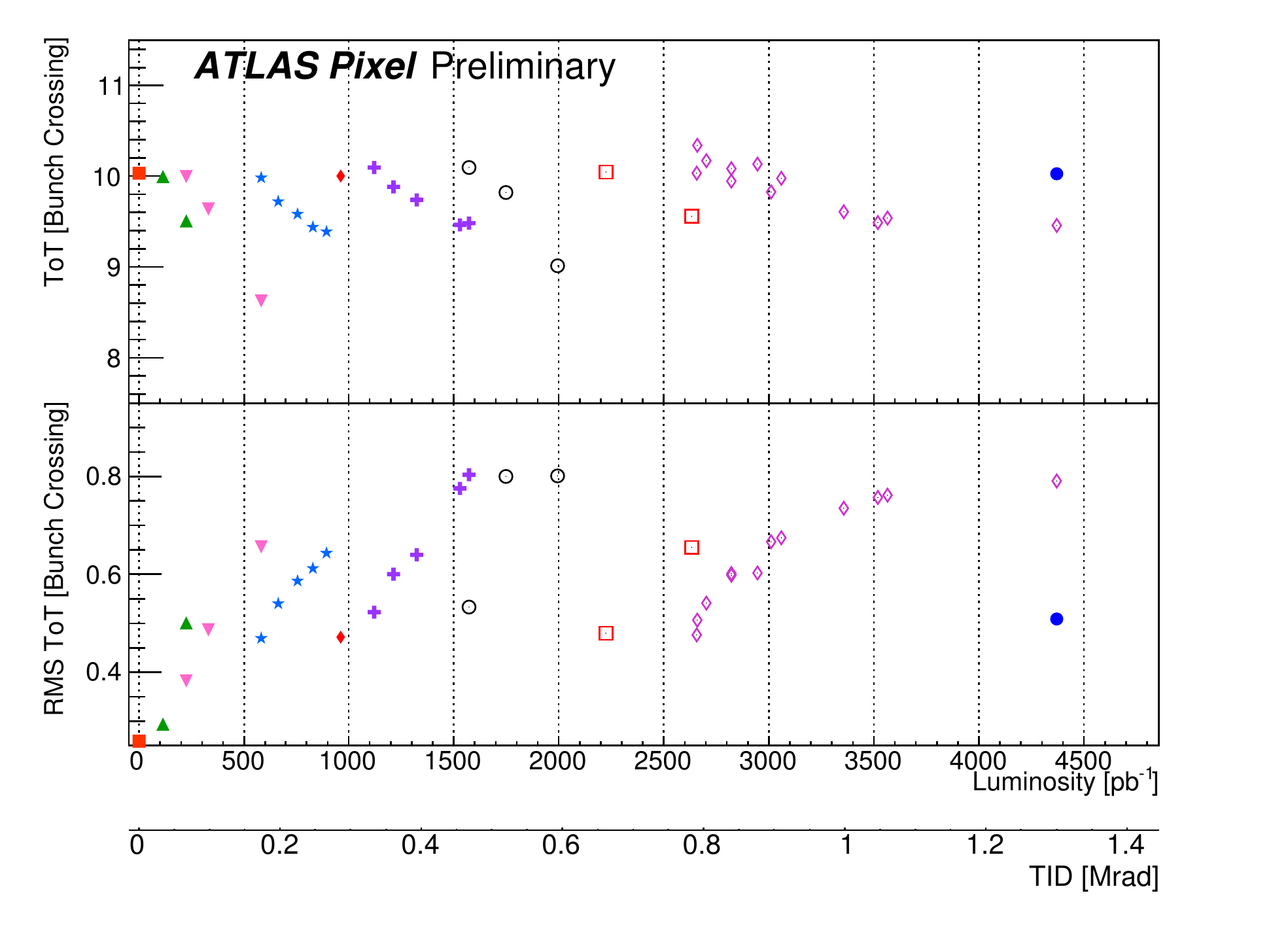}
\caption{The time-over-threshold (ToT) and its RMS as a function of the integrated luminosity or total ionising dose (TID)~\cite{IBLdetuning}. The detector was regularly retuned, and each marker type corresponds to a single tuning of the detector.}
\label{fig:IBL_ToTdrift}
\end{figure}

In addition, as shown for example in Figure~\ref{fig:IBL_ToTdrift}, the calibration of the FE-I4 chips for the analog discriminator threshold and the target ToT were observed to drift rapidly despite frequent updating of the calibration.

\section{Irradiation test-results}

The increase of the LV current  of the FE-I4 chip and the drifting of its tuning parameters were traced back to the generation of a leakage current in NMOS transistors induced by radiation higher than usual. The radiation induces positive charges that are quickly trapped into the shallow-trench-insolation (STI) oxide at the edge of the transistor. Their accumulation builds up an electric field sufficient to open a source-drain channel where the leakage current flows. If the accumulation of positive charges is relatively fast, the formation of interface states is a slower process. The negative charges trapped into interface states start to compete with the oxide-trapped charges with a delay. This is what gives origin to the so called rebound effect~\cite{FACCIO}. Dedicated laboratory measurements~\cite{LAURA} of irradiated single transistors in 130\,nm CMOS commercial technologies showed that the increase of the leakage current reaches its peak value between 1\,Mrad and 3\,Mrad. For higher TID the current decreases to a value close to the pre-irradiated one. 

To reproduce and analyse the effects described above during the FE-I4 chip operation, several  irradiations and electrical tests 
were performed by using  X-ray (Seifert RP149~\cite{Seifert}, and XRAD-iR-160~\cite{PXI}) and proton (Bern Cyclotron~\cite{BernCyclotron}) sources. Since the current increase in NMOS transistors depends on dose rate and temperature~\cite{FACCIO}, measurements under different temperature and dose rate conditions have been carried out to qualify this dependency.

\begin{figure}[h!]
\centering
\includegraphics[width=3.5in]{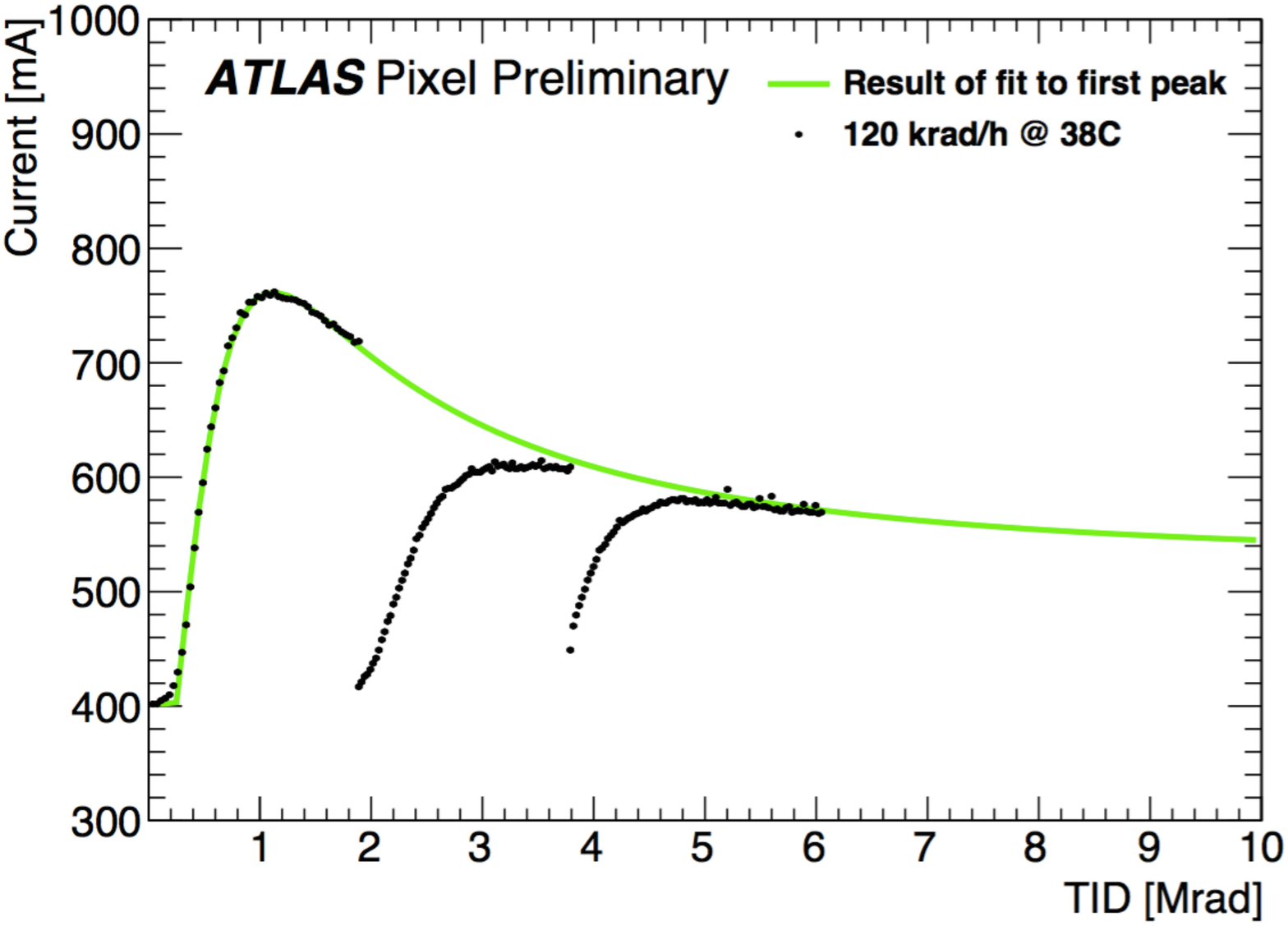}
\caption{Increase of the current consumption of a single FE-I4 chip in data taking condition as a function of the total ionising dose (TID). The temperature of the chip was 38\,$^\circ$C and the dose rate 120\,krad\,h$^{-1}$. After reaching the maximum of each peak the chip was annealed several hours resulting in the observed partial recovery~\cite{LVCurrentIrrad}. The fit performed on the first set of data (first peak) has been carried out by using the current parametrisation described in Ref.~\cite{MALTE}.}
\label{fig:ThreePeaks}
\end{figure}

The first irradiation test aimed at measuring the boundary current (at a given temperature and dose rate) that the chip always approaches after annealing periods and re-irradiation. Figure~\ref{fig:ThreePeaks} shows the increase of the current consumption of a single FE-I4 chip in data taking condition as a function of the TID. The temperature of the chip was 38\,$^\circ$C and the dose rate 120\,krad\,h$^{-1}$. After reaching the maximum of each peak the chip was annealed for several hours resulting in the observed partial recovery. 

\begin{figure}[h!]
\centering
\includegraphics[width=3.5in]{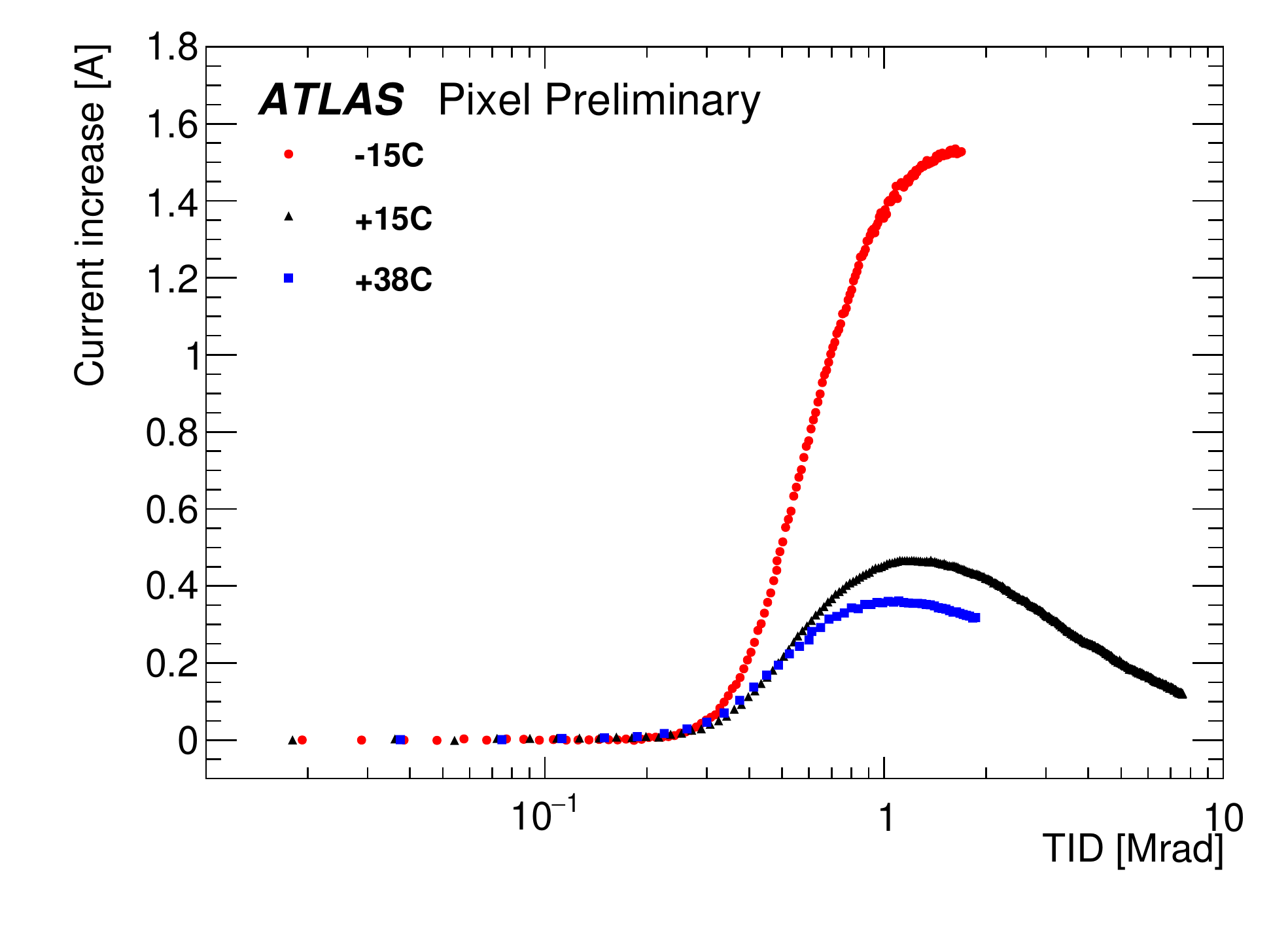}
\caption{Increase of the LV current of three single FE-I4 chips in data taking condition as a function of the total ionising dose (TID) 
in logarithmic x-axis scale. Test measurements were carried out at 38\,$^\circ$C (blue points), at 15\,$^\circ$C (black points) and at $-$\,15\,$^\circ$C (red points) with a dose rate of 120\,krad\,h$^{-1}$. A dose rate up to 10\,krad\,h$^{-1}$ is expected in the experiment. The LV current of the single FE-I4 chips before irradiation were 400\,mA (38\,$^\circ$C), 360\,mA (15\,$^\circ$C) and 380\,mA ($-$\,38\,$^\circ$C)~\cite{LVCurrentIrrad}.}
\label{fig:TemperatureComparison}
\end{figure}

\begin{figure}[h!]
\centering
\includegraphics[width=3.5in]{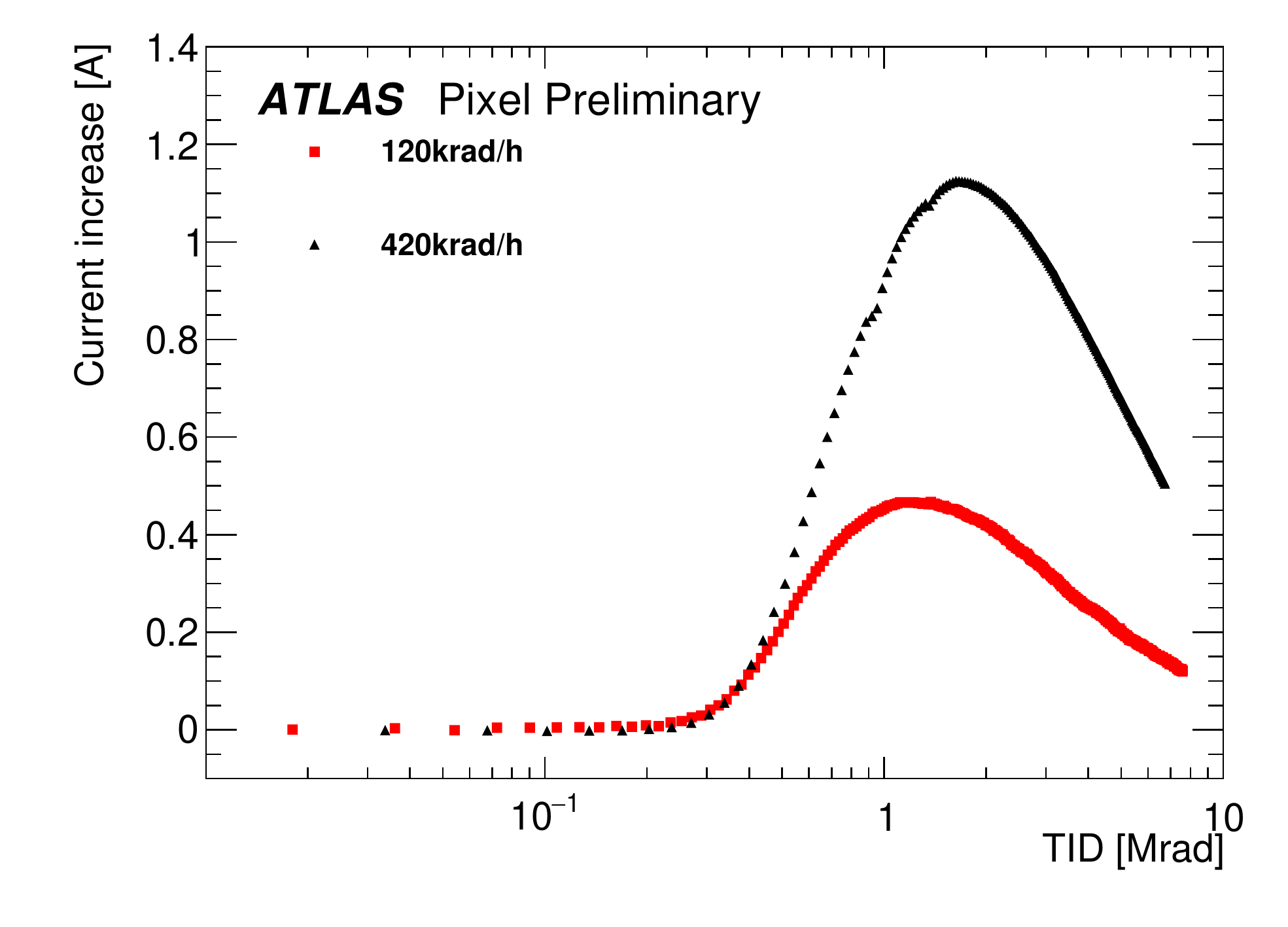}
\caption{Increase of the LV current  of two single FE-I4 chips in data taking condition as a function of the total ionising dose (TID) in logarithmic x-axis scale. Test measurements were carried out at 15\,$^\circ$C with a dose rate of 120\,krad\,h$^{-1}$ (red points) and 420\,krad\,h$^{-1}$ (black points). A dose rate up to 10\,krad\,h$^{-1}$ is expected in the experiment. The LV current of the single FE-I4 chips before irradiation were 380\,mA (420\,krad\,h$^{-1}$) and 360\,mA (120\,krad\,h$^{-1}$)~\cite{LVCurrentIrrad}.}
\label{fig:DoserateComparison}
\end{figure}

Then, to study the dependence of the LV current increase on temperature and dose rate several irradiation tests were performed by setting one of those  variables and changing the other. 
Figure~\ref{fig:TemperatureComparison} shows the results of three different measurements, performed with 
three different and previously not irradiated chips. The dose rate was 120\,krad\,h$^{-1}$  and the temperatures were 38\,$^\circ$C, 
15\,$^\circ$C and $-$\,38\,$^\circ$C. Before irradiation the LV current of the three chips was 400\,mA (38\,$^\circ$C), 
360\,mA (15\,$^\circ$C) and 380\,mA ($-$\,38\,$^\circ$C).
For comparison Figure~\ref{fig:DoserateComparison} shows the result of two different measurements where the temperature was kept fix at 15\,$^\circ$C, while the dose rate  set to 120\,krad\,h$^{-1}$  or 420\,krad\,h$^{-1}$. Also in this case the tests were performed with different and previously not irradiated chips. 

The measurements described above revealed two facts: i) at a given dose rate the LV current increase is stronger at lower temperatures; ii) at a given temperature, the LV current increase is stronger at higher dose rates.

\begin{figure}[h!]
\centering
\includegraphics[width=3.5in]{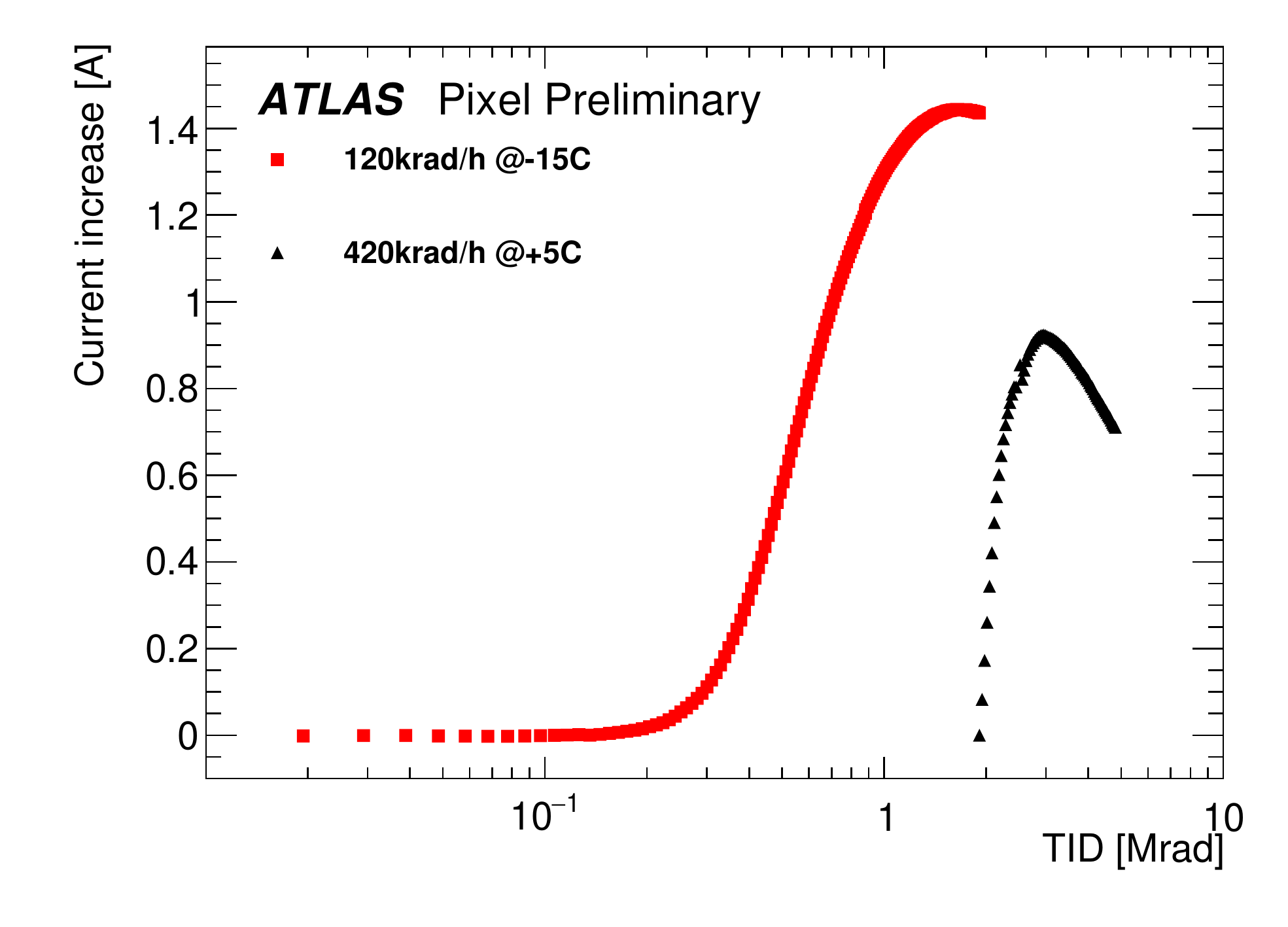}
\caption{Increase of the LV current of a single FE-I4 chip in data taking condition as a function of the total ionising dose (TID) in logarithmic x-axis scale during two consecutive irradiation campaigns in a lab measurement. Between the two irradiations several hours of annealing period at room temperature was performed, and resulted in the observed recovery. The TID of both irradiations is summed up. The LV current of a single FE-I4 chip before irradiation was 380\,mA (first step) and 550\,mA (second step)~\cite{LVCurrentIrrad}.}
\label{fig:MixedIrradiaiton}
\end{figure}

To simulate the dose rate conditions of the 2015 and 2016 data taking, a first irradiation was performed at $-$15\,$^\circ$C and 120\,krad\,h$^{-1}$. This was followed by several hours of annealing and a second irradiation this time performed at 5\,$^\circ$C and 420\,krad\,h$^{-1}$.
As shown in Figure~\ref{fig:MixedIrradiaiton} the second LV current peak is lower than the first one, i.e. by increasing the operational temperature of the chip it was possible to keep the increase of the LV current below the boundary current given by the first irradiation. 

To verify that a temperature of 5\,$^\circ$C is safe for the IBL detector operation, a measurement  at 10\,krad\,h$^{-1}$ was performed. The maximum LV current increase was of the order of 250\,mA, which gives a LV current increase of 1\,A for a four-chip unit, which would not exceed the safety limit of the LV current originally set to 2.8\,A.

In principle, lower operational temperatures are favourable for the sensor performance and properties after irradiation and therefore preferred. Consequently, irradiation and electrical tests were also performed at a temperature of 0\,$^\circ$C to investigate the feasibility for a colder operation.
In addition investigated was the evolution of the maximum of the LV current peak under several irradiation steps followed, interleaved with periods of annealing.
In this case the first two consecutive peaks of the LV current increase exceeded the maximum current allowed for a safe detector operation. Therefore, it was decided to set 5\,$^\circ$C as minimum temperature for a safe and successful data taking.

\section{Detector operation guideline}

Based on the observations during the first year of data- taking in 2015 with the IBL detector, it was decided to raise the safety limit for the IBL LV currents from 2.8\,A to 3\,A for module groups of four chips, which means a current consumption of 750\,${\mathrm{mA}}$ per chip. 
Since the average current consumption for a sigle FE-I4 chip  is about 400\,${\mathrm{mA}}$ before irradiation, the increase of the current due to the TID effects can not be higher than 350\,${\mathrm{mA}}$ per chip.

Given the above results it was decided to increase the IBL operation temperature from $-$\,10\,$^\circ$C to 15\,$^\circ$C. 
In addition,  the digital supply voltage (V$_{\mathrm{D}}$) was lowered from 1.2\,V to 1\,V to decrease the LV current. 

Thanks to dedicated measurements at 5\,$^\circ$C and at a dose rate comparable to the LHC in 2016 (10\,krad\,h$^{-1}$), it is proven that the current increase is of the order of 250\,${\mathrm{mA}}$. With this a module group of four chips does not exceed the safety limit of 3\,A. Therefore operating the IBL detector at 5\,$^\circ$C is safe with respect to the expected luminosity in 2016. 
The temperature of the IBL cooling system was lowered to a set point of 5\,$^\circ$C. The digital  supply voltage (V$_{\mathrm{D}}$) was raised from 1\,V to 1.2\,V, after an accumulated dose of $\sim$5\,Mrad which, as the measurements show, is well beyond the high peak region for the current consumption. 

\begin{figure}[h!]
\centering
\includegraphics[width=3.5in]{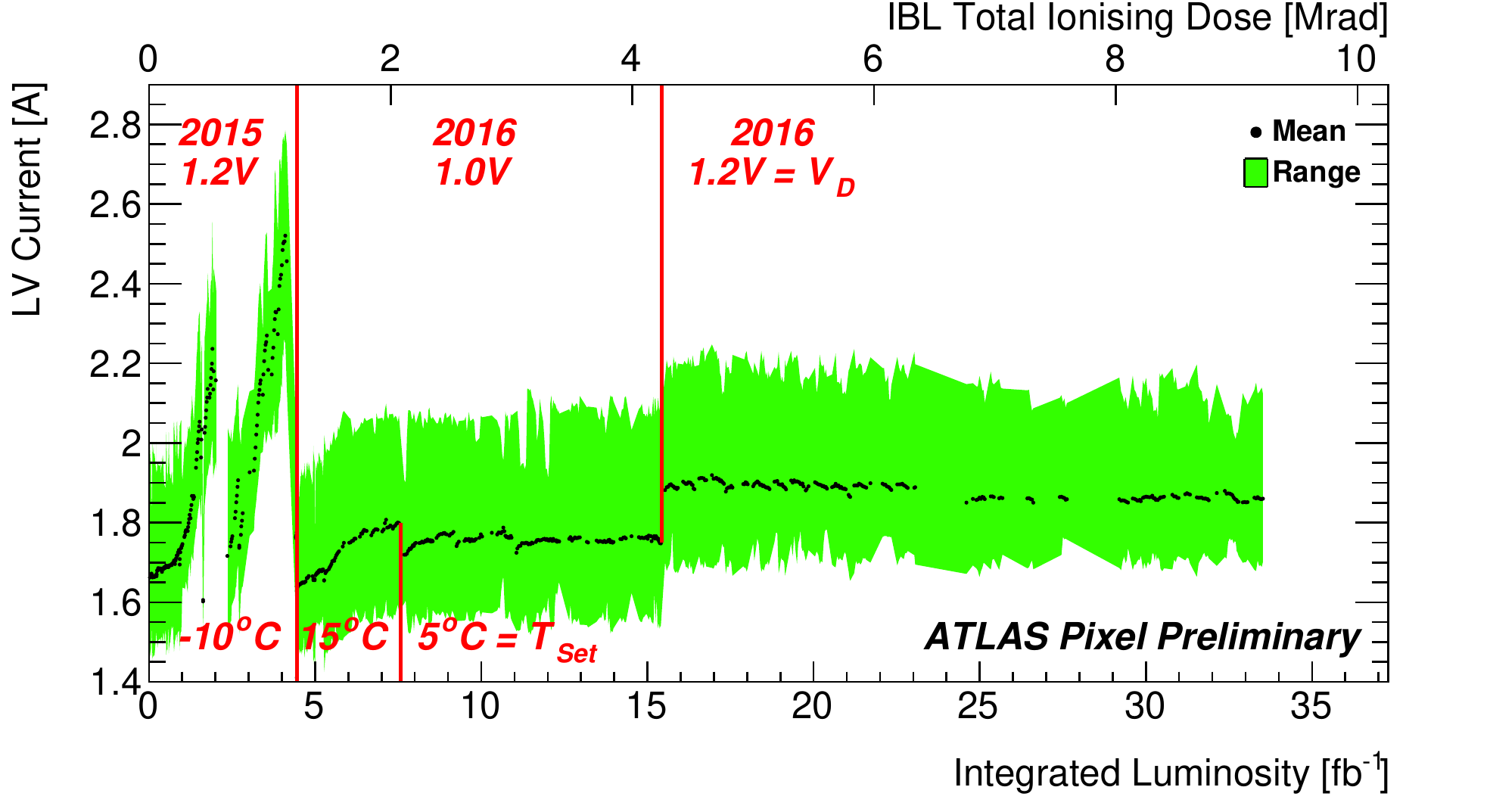}
\caption{Mean Low Voltage (LV) current in IBL FE-I4 chips during stable beam against integrated luminosity and total ionising dose (TID); values are averaged for all modules across 100 luminosity blocks. Changes in digital voltage (V$_{\mathrm{D}}$) are highlighted. The set temperatures (T$_{\mathrm{Set}}$) of the modules correspond to actual module temperatures of about -5$^\circ$C, 20$^\circ$C and 10$^\circ$C. There were significant increases in LV current during 2015; this was addressed in 2016 by increasing the module temperatures and decreasing the digital voltage. The digital voltages were later increased back to decrease readout error frequency~\cite{IBLLVcurrent}.}
\label{fig:LV_vs_Lint_vs_TID}
\end{figure} 

An overview of the mean LV current of the IBL FE-I4 chips as a function of integrated luminosity and TID during stable beam is shown in Figure~\ref{fig:LV_vs_Lint_vs_TID}. The LV currents are averaged for all modules across 100 luminosity blocks, and the changes in digital supply voltage (V$_{\mathrm{D}}$) and the temperature (T$_{\mathrm{Set}}$) are highlighted. 

In addition, since the shift of the tuning parameters can be seen even at low dose rates and warmer temperatures, a retuning on a regular basis was performed.

\section{Summary}

The Insertable B-Layer (IBL) is the innermost pixel barrel layer of the ATLAS detector installed in 2014. 

During the first year of data taking in 2015, a peculiar increase of the LV current of the front-end chip and the detuning of its parameters (threshold and time-over-threshold) have been observed in relation to received total ionising dose. It was tracked back to the generation of a leakage current in NMOS transistors induced by radiation.

Dedicated irradiation and electrical tests of FE-I4 chips showed that the leakage current reaches its peak value when the total ionising dose is in the range of 1\,Mrad -- 3\,Mrad, and above this the current decreases to a value close to the pre-irradiation one. This effect was shown to be temperature and dose rate dependent.

Thanks to intensive studies it was possible to apply special detector settings to still guarantee a successful data-taking.

\end{document}